\documentstyle[floats,aps]{revtex}

\begin{document}

\tighten

\title{\bf PHYSICS OF $Q^2$-DEPENDENCE IN THE NUCLEON'S \\
$G_1(x,Q^2)$ STRUCTURE FUNCTION SUM RULE}

\author{Xiangdong Ji\thanks
{\baselineskip=13pt
This work is supported in part by funds provided by the U.S.
Department of Energy (D.O.E.) under cooperative agreement
\#DF-FC02-94ER40818.}%
\medskip
\thanks{\baselineskip=13pt
Talk presented at the Workshop on Spin Degrees of Freedom
in Electromagnetic Nuclear Physics
APS/DNP fall meeting,
Williamsburg, Virginia, 1994.} \\
{\em Center for Theoretical Physics \\
Laboratory for Nuclear Science \\
and Department of Physics \\
Massachusetts Institute of Technology \\
Cambridge, Massachusetts~ 02139~~U.S.A.} \\
{~}}

\setlength{\baselineskip}{2.6ex}

\date{MIT-CTP-2411 \hfill  HEP-PH/9502288
 \hfill January 1995}

\maketitle

\begin{center}
\parbox{13.0cm}
{\begin{center} ABSTRACT \end{center}
{\small \hspace*{0.3cm}
I discuss in this talk the physics of the $Q^2$ dependence
of the $G_1(x,Q^2)$ structure function sum rule. For $Q^2>3$ GeV$^2$, the
$Q^2$ variation is controlled by pure QCD radiative corrections.
For $0.5<Q^2<3$ GeV$^2$, the twist-four contribution
becomes significant, but stays perturbative. For $Q^2$ below
$\sim 0.05$, the sum rule is determined by low-energy
theorems. The rapid change of the sum rule
between 0.05 and 0.5 GeV$^2$ signals the transition between
parton and hadron degrees of freedom.
}}
\end{center}

\section*{}


In polarized electron or muon scattering on a polarized nucleon
target, one measures the nucleon tensor,
\begin{eqnarray}
      W_{\mu\nu} &= &{1\over 2} \sum_n (2\pi)^3\delta^4(P+q-P_n) \nonumber \\
    && {\quad} \times \langle PS|J_\mu(0)|n\rangle
           \langle n|J_\nu(0)|PS\rangle \  ,
\end{eqnarray}
where $|PS\rangle $ is the ground state of the nucleon with momentum
$P^\mu$ and polarization $S^\mu$, $|n\rangle$ are the excited states of
the nucleon after absorbing the virtual photon of momentum $q^\mu$,
and $J_\mu$ is the usual electromagnetic current of the nucleon,
which are composed of quark fields.  The spin-dependent part
of the tensor is known to depend on two Lorentz scalar structure
functions $G_1(Q^2,\nu)$ and $G_2(Q^2, \nu)$,
\begin{equation}
      W_{\mu\nu}|_{\rm spin-depen.} = -i\epsilon_{\mu\nu\alpha\beta}
           q^\alpha \left({G_1\over M^2}S^\beta + {G_2\over M^4}
            \left(S^\beta\nu M - P^\beta(S\cdot q)\right)\right)
\end{equation}
where $Q^2=-q^2$ and $\nu M = P\cdot q$.

Although we shall not always work in the Bjorken limit, it turns out
convenient to replace variable $\nu$ by $x$:
$x=Q^2/2M\nu$. The drawback of doing this is that the whole photo-production
region shrinks to a point $x=0$ and $Q^2=0$. However, for our purpose
it is not a problem. I assume from now on that $G_1(Q^2, x)$
is measured to a good precision in low and intermediate $Q^2$ regions.
This may turn out to be the biggest assumption of my talk. I certainly
hope this can be done in the future at CEBAF and other places.

Two interesting sum rules exist for $G_1$ at large and small
$Q^2$, respectively. The deep-inelastic sum rule
is defined at $Q^2\to \infty$ limit,
\begin{eqnarray}
          \Gamma &= &\lim_{Q^2\to \infty}\int^1_0
  g_1(x, Q^2) dx \ ,  \nonumber \\
        &=&{1\over 2}\sum_i e_i^2 \Delta q_i \ ,
\end{eqnarray}
where $g_1(x, Q^2) = (\nu/M) G_1(\nu, Q^2)$ is the scaling
function and $\Delta q_i $ is the axial charge for quark
flavor $i$, which is defined by,
\begin{equation}
       \langle PS|\bar \psi_i\gamma_\mu\gamma_5\psi_i|PS\rangle
    = 2\Delta q_i S_\mu \ .
\end{equation}
The Bjorken sum rule relates $\Gamma^p-\Gamma^n$ to the neutron
$\beta$-decay constant $g_A$~\cite{BJ} and
Ellis-Jaffe sum rules refer to a model prediction for $\Gamma^p$
and $\Gamma^n$ made by Ellis and Jaffe~\cite{EJ}.

The Drell-Hearn-Gerasimov (DHG) sum rule is a sum rule for $G_1(\nu, Q^2)$
at the real photon point $Q^2=0$~\cite{DHG}. For simplicity of
discussion, I view the sum rule as the limit of $Q^2\to 0$,
\begin{equation}
          \lim_{Q^2\to 0} \int^\infty_{\nu_{\rm in}} {d\nu\over \nu}
        G_1(\nu, Q^2) = -{1\over 4}\kappa^2\ ,
\end{equation}
where $\nu_{\rm in}$ is the inelastic threshold and $\kappa$
is the anomalous magnetic moment of the nucleon.  Using the scaling
function, I can write,
\begin{equation}
      \int^1_0 dx g_1(x, Q^2)^{\rm inelastic} = -{\kappa^2\over 8}
    {Q^2\over M^2}  + {\cal O}\left(\left({Q^2\over M^2}\right)^2\right) \  ,
\end{equation}
for small $Q^2$. The question I want to address below is what physics
controls the variation of the sum rule between the large
and small $Q^2$ limits.

First let me consider deep-inelastic sum rules at large but finite
$Q^2$. There are two types of QCD corrections to the $Q^2\to \infty$
limit. The first is the QCD radiative corrections shown in Fig.~1(a),
which take into account the effects of hard gluons in the hard
process. The second is the higher-twist corrections shown
in Fig.~1(b), which are
basically initial and final state interactions between the active
quark and the remnants of the target. For example, the Bjorken
sum rule with these corrections reads,
\begin{figure}
\centering
\vspace{2in}
\centering
\includegraphics{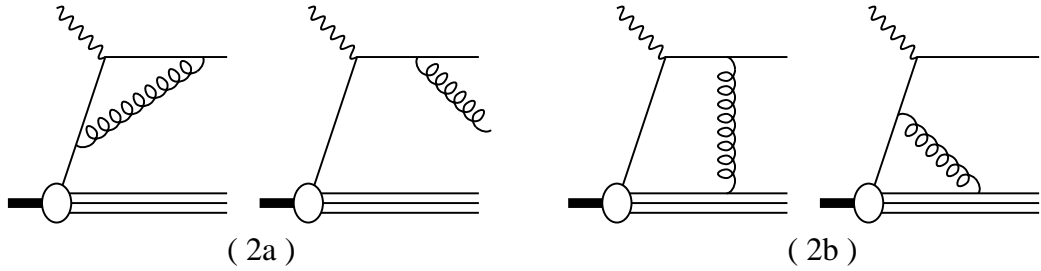}
\vspace*{-0.5in}   
\caption{%
a). QCD radiative corrections to deep-inelastic
scattering.
{}~b). Higher-twist corrections to deep-inelastic scattering.}
\label{fig1}
\end{figure}
\begin{equation}
        \int^1_0 g_1^{p-n}(x, Q^2) dx = {g_A\over 6}
           \left(1-{\alpha_s(Q^2)\over \pi}
          -\cdots \right)
           + {\mu_4^{p-n}(Q^2) \over Q^2} + \cdots
\label{BSR}
\end{equation}
where the terms in the bracket represent radiative effects and
$\mu_4^{p-n}$ is the nucleon matrix elements of some
twist-two, three, and four operators~\cite{JI1}.

A number of comments can be made about the sum rule in Eq.~(\ref{BSR}):
\begin{itemize}
\item{Theoretically there is an ambiguity in separating out contributions
of different twists. This was first recognized by
A. Mueller~\cite{MUL}. The problem is that the perturbative series for
radiative corrections is not convergent.  It is a non-Borel-summable
series. Thus the result obtained by Ellis and Karliner by comparing
the data with the four-loop prediction should be taken with a grain of
salt, particularly at low $Q^2$~\cite{EK}.  I have recently outlined a
solution to the problem~\cite{JI2}, but I cannot talk about it here
due to time limitation.  }
\item{The sum rule {\em must\/} include the elastic contribution
as it becomes important below $Q^2=2$ GeV$^2$. The reason is obvious:
the sum rule is derived from operator product expansion and one
gets an operator product only when all intermediate states are summed
over. If one is still not convinced, consider the nucleon is a
point-like particle, then the only contribution to the sum rule
is elastic scattering~\cite{JI3}. For the nucleon, the elastic contribution
to $\Gamma$ is,
\begin{equation}
    \Gamma^{\rm elastic} = {1\over 2}F_1(F_1+F_2) - {1\over 8M^2}F_2^2Q^2 \ .
\end{equation}
where $F_1$ and $F_2$ are the usual Dirac and Pauli form factors.}
\item{Higher-twist contributions have been estimated in the
MIT bag model~\cite{JI1} and
in QCD sum rule approach~\cite{BBK},}
\begin{eqnarray}
           \begin{array}{rcl@{\qquad}l}
               \mu^{p-n}_4 &=  & 0.031 M^2 \ ,     & \hbox{(Bag)} \\
               \mu^{p-n}_4 &= & -0.023 M^2  \ .    &  \hbox{(QSR)}
           \end{array}
\end{eqnarray}
Here the infrared renormalon problem has been ignored. The two estimates
differ in sign. This shows that we are not yet confident
in calculating higher-twist matrix elements. However, it is quite
clear that the size of the higher-twist contribution is small.
It contributes at 10\% level
at $Q^2=2$ GeV$^2$, and becomes negligible at $Q^2=10$ GeV$^2$.
\end{itemize}

The above discussion shows that {\em $\Gamma(Q^2)$ changes very little
from $Q^2=\infty$ down to $Q^2=0.5$ GeV\/}$^2$. Radiative effects are
on the order of 10\% to 20\% in the entire region. The twist-four
effects are important only in the range $Q^2\sim  0.5-3$ GeV$^2$.
Again, their contribution is not overwhelming.

Now let me turn to the sum rule at $Q^2\sim 0$. The DHG sum rule
certainly needs to be tested. Its validity tells us whether
there is a subtraction constant in the dispersion relation, whether
the sum rule is convergent, and
whether there are fixed pole contributions, etc. The detailed mechanism
for sum rule saturation is also interesting. In particular, there
are indications that the $\Delta$ excitation exhausts major part of
the sum rule.

One can generalizes the DHG sum rule to small $Q^2$ by writing
a low energy expansion,
\begin{equation}
     \int^1_0 dx g_1(x, Q^2)^{\rm inelastic}
    = -{\kappa^2\over 8}{Q^2 \over M^2}
           + \alpha \left({Q^2\over M^2}\right)^2+\cdots
\end{equation}
where $\alpha$ is a parameter which can be calculated for instance
in chiral perturbation theory. It is also interesting to
test this type of generalized sum rule.

One interesting question is how to connect the DHG sum rule to
the deep-inelastic sum rule. This question is first studied
by Anselmino, Ioffe, and Leader~\cite{AIL}, and the result has been quoted
by many authors. Unfortunately, their study is wrong.
They neglected the elastic contribution when interpolating
high and low $Q^2$ sum rules and thus got the incorrect
conclusion that
$\Gamma_p / Q^2$ has to change sign at some intermediate $Q^2$.
The sign change was considered as mysterious. As I said before,
as $Q^2$ decreases, the high $Q^2$ side physics is controlled by
twist expansion which, by definition, contains the elastic
contribution.

Once the elastic contribution is included, we have at low
$Q^2$,
\begin{eqnarray}
      \Gamma_p(Q^2)& =& \Gamma_p(Q^2)^{\rm elastic}
           + \Gamma_p(Q^2)^{\rm inelastic}\nonumber
\\
    &=& 1.396 - 8.631Q^2 + \alpha Q^4 + \cdots
\end{eqnarray}
According to the above, $\Gamma_p(Q^2\to 0 ) \to 1.396$, which is much
larger than $\Gamma_p(Q^2=10 \hbox{ GeV}^2 )\linebreak[4]
=0.136$. Clearly,
$\Gamma_p(Q^2)$ drops very quickly as $Q^2$ increases due to the large
coefficient of the $Q^2$ term. In fact, if one neglects
the higher order terms, $\Gamma_p(Q^2)$ drops to the level at
$Q^2 = 10$ GeV$^2$ when $ Q^2 = 0.15$ GeV$^2$. This behavior
is certainly consistent with the small higher-twist effects at
moderate $Q^2$. However, beyond that, the $Q^2\sim 0$ behavior
says nothing about the size of higher twist effects, contrary
to many claims in the literature.

Certainly, the change between $Q^2\sim 0.05$ and $Q^2=0.5$ is
interesting. We do not have reliable theoretical prediction in the
region. However, we believe that the transition between hadronic
and partonic description of scattering occurs in this region.
Since the transition is likely smooth, I think nothing
drastic happens for $\Gamma_p(Q^2)$ other than
a smooth connection between low and high $Q^2$ limits.

To summarize the above discussion, the physics of
$Q^2$ variation of $\Gamma_p(Q^2)$ can be roughly divided into
four regions. For $Q^2>3$ GeV$^2$, the $Q^2$ variation is
controlled by QCD radiative corrections. In $0.5<Q^2<3$ GeV$^2$,
the twist-four contribution is important. Below $\sim 0.05$ GeV$^2$,
the $Q^2$ variation is determined by low energy theorems. In
$0.05<Q^2<0.5$ GeV$^2$, parton-hadron transition happens.
A rough sketch for the sum rule variation (the solid line) is shown
in Fig.~2. The dotted line represents an extrapolation from high
energy and the dash-dotted line an extrapolation from low energy.

\begin{figure}
\centering
\vspace*{3in}
\includegraphics{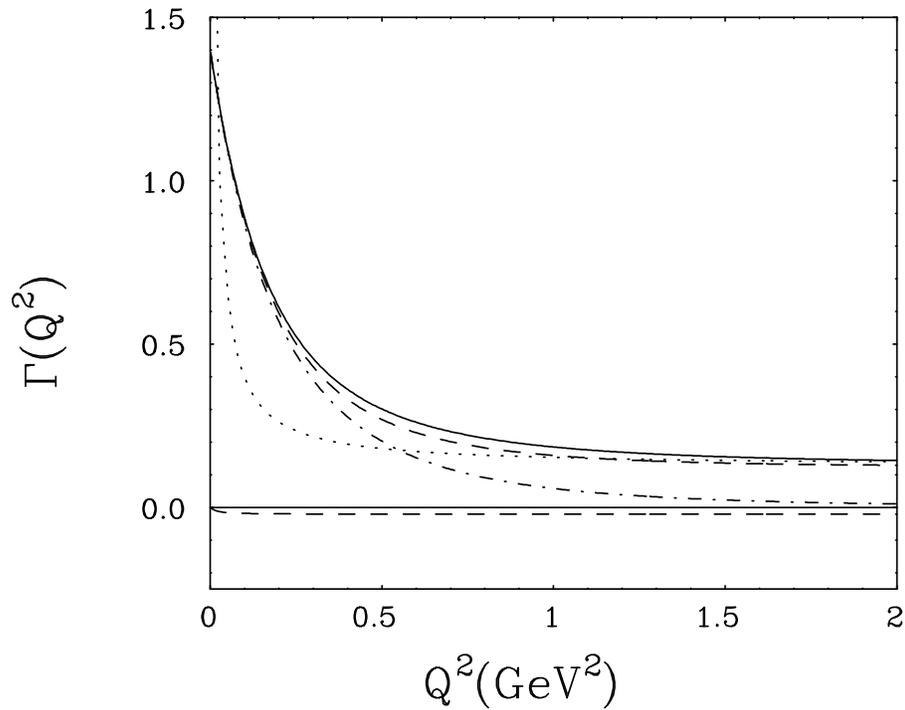}
\vspace*{0.8in}   
\caption{Schematic  $Q^2$-dependent of the first
moment of $G_1$ structure function.}
\label{fig2}
\end{figure}

Let me emphasize the importance of getting data
in the region $0.5<Q^2<3$ GeV$^2$. The sum rule here
can be constructed with future resonance data from CEBAF and
low $x$ data from SLAC or HERMES\@. It allows one
to extract the matrix element~$f$,
\begin{equation}
     \langle PS| g\bar \psi\tilde F^{\mu\nu}\gamma_\nu \psi|PS\rangle
   = 2fS^\mu \ .
\end{equation}
$f$ is very interesting from the nucleon's structure point of
view. In fact, the sign of $f$ determines roughly whether
the color magnetic field ${\bf B}$ in the polarized nucleon
is pointing to the direction of the spin or opposite.

Thus one can learn a lot of physics by measuring
the first moment of the $G_1$ structure function
at low and intermediate $Q^2$.
I urge experimenters go ahead to take some good data
on $G_1$.

\section*{}
\bibliographystyle{unsrt}

\end{document}